# Know Abnormal, Find Evil: Frequent Pattern Mining for Ransomware Threat Hunting and Intelligence

Sajad Homayoun, Ali Dehghantanha, Marzieh Ahmadzadeh, Sattar Hashemi, Raouf Khayami

*Abstract*—Emergence of crypto-ransomware has significantly changed the cyber threat landscape. A crypto ransomware removes data custodian access by encrypting valuable data on victims' computers and requests a ransom payment to re-instantiate custodian access by decrypting data. Timely detection of ransomware very much depends on how quickly and accurately system logs can be mined to hunt abnormalities and stop the evil. In this paper we first setup an environment to collect activity logs of 517 *Locky* ransomware samples, 535 *Cerber* ransomware samples and 572 samples of *TeslaCrypt* ransomware. We utilize Sequential Pattern Mining to find Maximal Sequential Patterns (MSP) of activities within different ransomware families as candidate features for classification using J48, Random Forest, Bagging and MLP algorithms. We could achieve 99% accuracy in detecting ransomware instances from goodware samples and 96.5% accuracy in detecting family of a given ransomware sample. Our results indicate usefulness and practicality of applying pattern mining techniques in detection of good features for ransomware hunting. Moreover, we showed existence of distinctive frequent patterns within different ransomware families which can be used for identification of a ransomware sample family for building intelligence about threat actors and threat profile of a given target.

*Index Terms*—Malware, ransomware, crypto ransomware, ransomware detection, ransomware family detection.

## I. INTRODUCTION

CYBERCRIMINALS pose a real and persistent threat to business, government and financial institutions all around the globe [1]–[3]. The volume, scope and cost of cybercrime all remain on an upward trend [4]. Malicious programs have always been an important tool in cyber criminals portfolios [5], [6] and almost everyday we are detecting new variants of malware programs [7]. Development and wide adoption of e-currencies such as Bitcoin led to many changes in cybercriminal activities [8], [9] including development of a new type of malware called ransomware [10]. Ransomware is a type of malware that removes a custodian access to her data and request for a ransom payment to re-instantiate data access [11]. Ransomware has been around since 1989, when ransomware first appeared under the name of AIDS Trojan [12]. Ransomwares are utilizing different infection vectors ranging from social engineering and Spam emails to botnets for distribution.

There are two main types of ransomwares namely *Locker* and *Crypto* ransomwares. The former locks a system and denies users' access without making any changes to the data stored on the system while the latter encrypts all or selected data usually using a strong cryptography algorithm such as AES or RSA [12]. After encryption, the victim is presented with the ransom payment instructions with possibility of recovering ransomed data.

Ransomware has dominated the threat landscape in 2016 with annual increase rate of 267% [13]. It is estimated that in 2014 only, cybercriminals have made more than $3 million profit using ransomware programs [14]. Unsurprisingly, ransomware attacks are mostly infecting individuals who are not security aware. Taking regular backups to a secure location is a good counter measure to reduce effects of ransomware infection [15]. These days, ransomware programs are indiscriminately targeting all industries ranging from healthcare to the banking sector and even power grids [4]. The Crypto-ransomware programs are much more popular than Lockers as almost always security engineers could find ways to unlock a system without paying the ransom while the only viable solution for decrypting strongly encrypted data is to pay ransom and receive decryption key [16]. Therefore, focus of this paper is only on crypto-ransomware and in the rest of the paper, the word "ransomware" is actually referring to the "crypto-ransomware" only. It was already reported that cyber security training and employee awareness would reduce the risk of ransomware attacks [17]. However, automated tools and techniques are required to detect ransomware applications before they are launched [18] or within a short period after their execution [19]. The growing danger of ransomware attacks requires new solutions for prevention, detection and removing ransomwares programs.

In this paper, we are using a sequential pattern mining technique to detect best features for classification of ransomware applications from benign apps as well as identifying a ransomware sample family. We investigate usefulness of our detected features by applying them in *J48, Random Forest, Bagging* and *MLP* classification algorithms against a dataset contains 517 *Locky* ransomware samples, 535 *Cerber* ransomware samples, 572 samples of *TeslaCrypt* ransomware and 220 standalone Windows Portable and Executable (PE32) benign applications. We not only achieved 99% accuracy in detection of ransomware samples and 96.5% in detection of

S. Homayoun, M. Ahmadzadeh and Raouf Khayami are with the Department of Computer Engineering and Information Technology, Shiraz University of Technology, Shiraz, Iran. e-mail: S.Homayoun@sutech.ac.ir.

A. Dehghantanha is with Department of Computer Science, School of Computing, Science and Engineering, University of Salford, Salford, U.K.

S. Hashemi is with Department of Computer Engineering, Shiraz University, Shiraz, Iran.

their families but reduced the detection time to less than 10 seconds of launching a ransom application; a third of the time reported by earlier studies i.e. [20]. Our results are not only indicative of usefulness of pattern mining techniques in identification of best features for hunting ransomware applications but show how patterns of different ransomware families can help in detecting a ransomware family which assist in building intelligence about threats applicable to a given target. To the best of authors knowledge this is the very first paper applying sequence pattern mining to detect frequent features of ransomware applications and to build vectored datasets of ransomware applications logs. Our created datasets contain logs of Dynamic Link Libraries (DLL) activities, file system activities and registry activities of 1624 ransomware samples from three different families and 220 benign applications.

We are using widely accepted criteria namely True Positive (TP), False Positive (FP), True Negative (TN), and False Negative to evaluate our model [21]–[23]. TP is reflecting total samples that correctly identified. FP shows incorrectly identified samples. TN demonstrates the number of correctly rejected samples, while FN shows incorrectly rejected samples. *Precisions* of a classification algorithm is a measure of relevancy of results and is calculated by dividing TP by total of FP and TP predicted by a classifier as shown in equation (1). *Recall* reflects the proportion of positives that are correctly identified by classification technique which is calculated by dividing TP by total of TP and FN as shown in equation (2). F-measure is showing the performance of a classification algorithm and is calculated by the harmonic mean of precision and recall as shown in equation (3).

$$Precision = \frac{TP}{TP+FP} \quad (1)$$

$$Recall = \frac{TP}{TP+FN} \quad (2)$$

$$F-measure = 2 \times \frac{Precision \times Recall}{Precision + Recall} \quad (3)$$

We will also report *Receiver Operating Characteristic (ROC)* that is a potentially powerful metric for comparison of different classifiers, because it is invariant against skewness of classes in the dataset. In a *ROC* curve the true positive rate is plotted in function of the false positive rate for different thresholds. In addition to *ROC*, *Area Under the Curve (AUC)* is a measure of how well a parameter can be used to distinguish between two classes. *AUC* is a single value that summarizes the *ROC* by calculating the area of the convex shape below the *ROC* curve. *AUC* can be between 0 and 1, where the value of 1 shows optimal point of perfect prediction.

*Matthews Correlation Coefficient (MCC)* [24] provides another measures of quality to compare different classifiers [25]. The *MCC* value is between −1 and +1, where in cases of perfect prediction it gives +1. −1 coefficient shows total disagreement between prediction and observation while the coefficient value of 0 indicates that the classifier does not work better than a random prediction. *MCC* is also a useful measure of classifier performance against imbalanced datasets. While *Precision, Recall or F-measure* values in a random guessing would be higher than *0.5*, *MCC* value would be around 0 for random guessing. Therefore, for making sure that our classifiers are far from random classifiers, we will compute *MCC* values for each classifier. The values can be computed using equation (4).

$$MCC = \frac{TP \times TN - FP \times FN}{\sqrt{(TP+FP)(TP+FN)(TN+FP)(TN+FN)}} \quad (4)$$

The remainder of this paper is organized as follows. Section II reviews some related research while Section III explains our method for collecting and preprocessing of data in a controlled environment. We describe feature extraction and vectorization in Section IV. Section V introduces our approach for ransomware detection followed by Section VI that describes our performance in detecting ransomwares families. Finally, section VII discusses about the achievements of this paper and concludes the paper.

## II. RELATED WORK

Ransomware programs are reportedly becoming a dominant tool for cybercriminals and a growing threat to our ICT infrastructure [10], [26], [27]. The possibility of using encryption techniques to encrypt users' data as part of a Denial of Service (DoS) attack is known for a very long time [28]. However, recent adoption of eCurrencies such as BitCoin provided many new opportunities for attackers including receiving a ransom payment for decrypting users' data [28]. In spite of its simplicity and primitive utilization of cryptographic techniques [29], ransomware programs are becoming a major tool in cyber criminals' toolset [30]. For any cyber threat, prevention is ideal but detection is a must and ransomware is not an exception [7], [31].

Situational cyber security awareness plays an important role in preventing cyber-attacks [32]. An educational framework that is tailored to ransomware threats [17] as well as a tool which mimicked ransomware attacks [33] proved to be useful in reducing ransomware infections. Moreover, technical countermeasures such verifying applications trustworthiness when calling a crypto library [34] or minimizing attack surface by limiting end-users' privilege proved effective in preventive ransomware attacks [16].

Most ransomwares detection solutions are relying on filesystem [35]–[37] and registry events [38] to identify malicious behaviors. Investigation of 1359 ransomware samples showed that majority of ransomware samples are using similar APIs and generating similar logs of filesystem activities [36]. For example, using 20 types of filesystem and registry events as features of a *Bayesian Network* model against 20 Windows ransomware samples resulted to an accurate ransomware detection with *F-Measure* of 0.93 [38]. UNVEIL [36] as a rasnsomware classification system utilized filesystem events to distinguish 13,637 ransomwares from a dataset of 148,223 malware samples with accuracy of 96.3%. *CloudRPS* [39] was a cloud-based ransomware detection system which relied on abnormal behaviors such as conversion of large quantities of files in a short interval to detect ransomware samples. *ElderRan* [20] utilized association between different operating system events to build a matrix of applications activities and to detect ransomware samples within 30 seconds of their execution with *AUC* of 0.995. Timely detection of a ransomware upon its execution is very crucial and systems that fail to detect ransomware in less than 10 seconds are not considered effective [11]. Moreover, timely identification of a ransomware family would assist in building intelligence about applicable threat actors and threat profile for a given target.

## III. Data Creation

We have downloaded 1624 Windows Portable Executable (PE32) ransomware samples from *virustotal.com* which were reported as malicious ransomware file by *Ransomware-Tracker.abuse.ch* in the period of February 2016 to March 2017. Collected samples belong to three families of ransomware namely 517 *Locky* samples, 535 *Cerber* samples and 572 samples of *TeslaCrypt*. The best type of goodware counterpart for malware applications are portable and standalone benign apps [32]. Therefore, we have collected all 220 available portable Windows PE32 benign applications from *portableapps.com*[1] in April 2017 to serve as goodware counterpart of our dataset.

We have setup the environment shown in Fig. 1 to collect logs of ransomware and goodware samples runtime activities. The Controller application on the host machine is randomly selecting a ransomware or goodware sample and passes it through FTP server to the Virtual Machine (VM). When the sample is successfully transferred, the Controller notifies the Launcher app to run the *ProcessMonitor* application and executes a given sample. Similar to the previous research [11], the first 10 seconds log of ransomware and benign applications runtime activities is collected and the created log file is uploaded to the Log repository on the host machine. Since majority of benign applications require human interactions to run (i.e clicking on a button), we have developed an application called *PyWinMonkey* which automates user interactions with an application. When the log file is successfully stored on the host machine, the Controller application reverts the VM back to its original copy and passes the next sample. It is notable that *PyWinMonkey* is similar to *Monkey*[2] Android app which utilized in many previous Android malware research papers [40] for mimicking human interactions. We have used Python 3.6.1 to develop Controller, Launcher and *PyWinMonkey* apps (accessible at https://github.com/sajadhomayoun/PyWinMonkey) and run *ProcessMonitor V3.31* on Windows10 build number 10240 on a computer with Core i7 CPU with 8 cores of 4GHz and 16GB of RAM. For each and every process, *ProcessMonitor* records loaded Dynamic Linked Libraries (DLLs), file system activities and registry activities. We scanned all captured logs to find unique activities throughout the dataset (see Table I). Therefore, we will have three sets of events namely $RegistryEvents_{Set}$, which includes all registry events, $DLLEvents_{Set}$, which includes all DLL events and $FileSystemEvents_{Set}$, which contains all Filesystem events as listed in Table I. Moreover, *EventType(E)* is a procedure that returns the type of given event (R for Registry events, F for Filesystem events, and D for DLL events) as shown in Fig. 2.

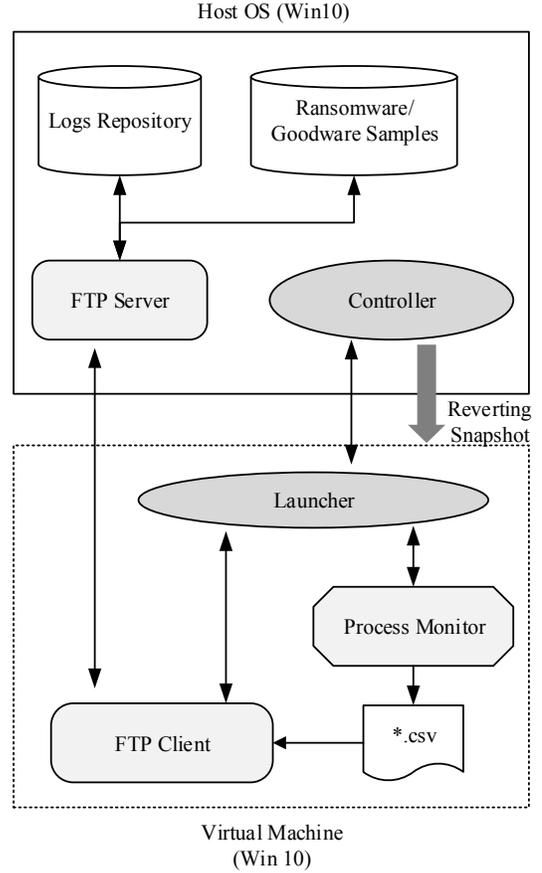

Fig. 1. Environment Setup to Capture Malware and Goodware Activities Log

As we will be using a sequential pattern mining technique (MG-FSM) to detect candidate features for classification task, we should convert our data into a sequential dataset which is a collection of sequences such as $D = \{S_1, S_2, ..., S_n\}$ where $S_i$ represents a sequentially ordered set of events. We have created a sequence of runtime events for each and every ransomware and benign application. $S_i$ represents a sequence of all events $E$ caused by launching an application $i$ ordered by time as follow:

$S_i = \{E_{1,i}(argE_1), E_{2,i}(argE_2), ..., E_{2,i}(argE_n)\}$ where $E_{x,y}(argE_x)$ represents event $x$ for an application $y$ and $argE_x$ shows the argument passed to the event $E_x$.

For example, $\{LoadImage(C : \backslash system32 \backslash gdi32.dll)\}$, $\{ReadFile(C : \backslash Windows \backslash SysWOW64 \backslash wininet.dll)\}$ shows a sequence of two events where the first event

---

1: **procedure** EventType(Event E)
2:    **if** $E \in RegistryEvents_{Set}$ **return R**
3:    **if** $E \in FilesystemEvents_{Set}$ **return F**
4:    **if** $E \in DLLEvents_{Set}$ **return D**
5: **end procedure**

Fig. 2. Determining Even Type of a given event.

---

[1] https://portableapps.com/apps
[2] https://developer.android.com/studio/test/monkey.html

TABLE I
LIST OF ACTIVITIES CAN BE CAPTURED BY PROCESS MONITOR

| Activity Type | List |
|---|---|
| Registry | RegQueryKey, RegOpenKey, RegQueryValue, RegCloseKey, RegCreateKey, RegSetInfoKey, RegEnumKey, RegQueryKeySecurity, RegEnumValue, RegSetValue, RegDeleteValue, RegQueryMultipleValueKey, RegDeleteKey, RegLoadKey, RegFlushKey |
| Filesystem | QueryNameInformationFile, ReadFile, CreateFile, QueryBasicInformationFile, CloseFile, QueryStandardInformationFile, CreateFileMapping, QuerySizeInformationVolume, FileSystemControl, QueryDirectory, WriteFile, QueryNetworkOpenInformationFile, QueryRemoteProtocolInformation, QuerySecurityFile, LockFile, UnlockFileSingle, DeviceIoControl, SetEndOfFileInformationFile, FlushBuffersFile, SetAllocationInformationFile, SetBasicInformationFile, QueryAttributeTagFile, QueryFileInternalInformationFile, QueryInformationVolume, QueryAttributeInformationVolume, SetRenameInformationFile, QueryNormalizedNameInformationFile, NotifyChangeDirectory, QueryFullSizeInformationVolume, SetSecurityFile, QueryStreamInformationFile, SetDispositionInformationFile, QueryEaInformationFile, QueryAllInformationFile, QueryIdInformation, SetPositionInformationFile, QueryPositionInformationFile, SetValidDataLengthInformationFile |
| DLL | LoadImage |

TABLE II
CREATED DATASETS

| Dataset | Number of Sequences |
|---|---|
| $D_{Locky}$ | 450 |
| $D_{Cerber}$ | 470 |
| $D_{TeslaCrypt}$ | 507 |
| $D_{Goodware}$ | 200 |
| $D_{OF}$ | 174 |

loads *gdi32.dll* in the memory of calling process (hence $C:\backslash system32\backslash gdi32.dll$ is the parameter for this event) and the second event reads *wininet.dll* file located at $C:\backslash Windows\backslash SysWOW64$. The size of each sequence depends on the number of events that are called by an application and varies between different apps.

Once all sequences are created, we have utilized the *Find Frequent Pattern Outlier Factor (FindFPOF)* [41] algorithm to remove any outlier sequence from our sequential dataset. It is notable that *FindPOF* is among very few sequential dataset outlier detection techniques which offers a reasonable detection performance [42]. *FindFPOF* benefits *Frequent Pattern Outlier Factor (FPOF)* to extract all frequent patterns from a dataset and removes outlier sequences as those with the least frequent patterns. Outliers were detected and removed for each ransomware family separately as it is expected that ransomware from the same family expose common features in compare with those from different families.

Table II reflects final datasets with the number of sequences in each dataset. We use $D_x$ notation to refer dataset $x$ in the rest of this paper. $D_{Locky}$ represents sequences of *Locky* ransomware samples, $D_{Cerber}$ shows *Cerber* ransomware sequences and $D_{TeslaCrypt}$ includes sequences of *TeslaCrypt* ransomware samples. $D_{Ransomware}$ represents combined sequences of all ransomware samples while $D_{Goodware}$ includes sequences of events of all benign applications. We randomly collected 52 *Locky*, 50 *Cerber*, 52 *TeslaCrypt* and 20 benign applications sequences in a separated dataset for over-fitting test as well ($D_{OF}$).

## IV. FEATURE EXTRACTION AND VECTORIZATION

To detect the best features for classification task, we need to first define detectable patterns of events and then utilize a pattern mining algorithm to find *Maximal Sequential Patterns (MSP)* collections within each dataset. Afterwards, every sequence within every relevant dataset is traversed based on a given *MSP* collection to provide features for training classifiers.

Sequential pattern mining techniques discover all subsequences (Sequential Patterns) that appear in a given sequential dataset with frequency of no less than a user-specified threshold ($min_{sup}$) [42]. A sequence $\alpha = \{a_1, a_2, ..., a_n\}$ is called a subsequence of another sequence $\beta = \{b_1, b_2, ..., b_m\}$ and $\beta$ is a super-sequence of $\alpha$, denoted as $\alpha \subseteq \beta$, if there exists integers $1 \leq j_1 < j_2 < ... < j_n \leq m$ such that $a_1 \subseteq b_{j_1}, a_2 \subseteq b_{j_2}, ..., a_n \subseteq b_{j_n}$. A sequence $\alpha$ is said to be frequent and called a *Sequential Pattern (SP)* in a sequential dataset $D$ if $sup_\alpha \geq min_{sup}$, where $sup_\alpha$ (support of $\alpha$) denotes the frequency of occurrence of $\alpha$ in a given sequential dataset $D$. Moreover, if a *Sequential Pattern SP* is not contained in any other sequential patterns, it is called a *Maximal Sequential Pattern (MSP)*. Collection of all *MSPs* with in a given sequential dataset $D$ can be denoted as a *Maximal Sequential Pattern Collection* ($MC_D$). Members of a *MC* are in format of $(P, sup_P)$ where $P$ is a *MSP* and $sup_P$ shows the frequency of occurrence of *P* in a given dataset $D$.

There are two major types of sequential pattern mining algorithms to extract *MSPs* namely Apriori-based and frequent pattern growth. Apriori-based algorithms are detecting *MSPs* based on the fact that any subset of a frequent pattern must be frequent. However, recursive nature of Apriori-based algorithms increases complexity and running time of the algorithm [43]. On the other side, frequent pattern growth algorithms are using divide-and-conquer techniques to narrow down the search space *MSPs*. Due to the large number of elements in each sequence (greater than 5000 elements in each sequence in this paper), traditional algorithms e.g. *Generalized Sequential Pattern (GSP)* [44] are inefficient. In other words, few of the commonly used sequential pattern mining algorithms are capable of producing maximal patterns in a reasonable time [45]. To detect *MSPs* in this study, we utilize a widely used frequent pattern growth algorithm [46] called *"Mind the Gap: Frequent Sequence Mining (MG-FSM)"* [47]. *MG-FSM* is a parallel processing solution based on *Map-Reduce* [48] which can be easily deployed on a cloud infrastructure to provide desired scalability. Low threshold values for minimum support may generate a huge set of *MSPs* that affects the computational feasibility of our vectorization model. On the other hand, very high values of minimum support may remove useful *MSPs* in detecting ransomwares from our datasets. Therefore, we decided to choose $min_{sup}$ of 50% to achieve a reasonable performance while covering sufficient number of *MSPs* in our dataset. Applying *MG-FSM* against our datasets generates four *MSP* collections namely $MC_{D_{Locky}}$, $MC_{D_{Cerber}}$, $MC_{D_{TeslaCrypt}}$ and $MC_{D_{Ransowmare}}$.

$MC_D = \{(P_x, sup_{P_x}) | sup_{P_x} \geq min_{sup} \land \forall P_x(\nexists P_y(P_x \subseteq P_y))\}$.

We can distinguish three types of atomic *MSPs* and six types of single step transition *MSPs* within our sequential datasets as shown in Table III. Atomic *MSPs* are representing continuous events of the same type i.e. the atomic *MSP* of $F$ represents continuous Filesystem events. Single step transitions *MSPs* are representing a transition from one atomic *MSP* to another. For example, *MSP* of "$RD$" represents a sequence of registry events ($R$ atomic *MSP*) followed by a sequence of DLL events ($D$ atomic MSP). Since we will have one feature in our vectored dataset for each transition; in cases of considering multi step transitions we will have more features for each vector. Having too many features makes a dataset sparse and difficult to find a separation hyperplane. This issue is referred as *curse of dimensionality issue* [44] which states that as the dimensionality increases, the volume of the space increases so fast that the available data become sparse. In this research, we can calculate total number of features in each vector using equation (5), where 3 is the number of considered activities ($F, R, D$) and $x$ is the desired steps in each transition. Consider $t = a, b, c$ as a 2 steps *MSP*, $a$ can be one of 3 possible activities ($F, R, D$) and as a constraint to make transition we have $b \neq a$ (2 possible activities for $b$) while for the next transition we must have $c \neq b$ (2 possible activities for $c$). Therefore, we will have $(3 \times 2^i)$ part of formula in equation 5. For $x = 3$ (single step, 2 steps and 3 steps transitions), we will have total of 45 features. Therefore, we decided to only consider single step transitions to avoid sparsity in extracted features.

$$TotalFeatures = 3 + \sum_{i=1}^{x}(3 \times 2^i) \quad (5)$$

A *MSP* $P = \{E_1, ..., E_n\}$ is atomic if $\forall_{E_x, E_y \in P \land E_x \neq E_y}(EventType(E_x) = EventType(E_y))$.

A *MSP* $P = \{E_1, ..., E_n\}$ is a single step transition if $\exists_{E_x, E_y \in P \land E_x \neq E_y}(EventType(E_x) \neq EventType(E_y))$.

We can define a set that contains all *MSP* types (*MSP* $Type_{Set}$) and a procedure (*MSPType(MSP P)*) in Fig. 3 that returns type of given sequence *S* as follow:

$MSPType_{Set} = \{R, F, D, RF, RD, FR, FD, DR, DF\}$.

*Support Ratio (SR)* of a *MSP* is a value in the range of [0,1] that shows the possibility of occurrence of the *MSP* in a given dataset of ransomware and is calculated by dividing frequency of occurrences of *MSP* ($sup_{MSP}$) by the total number of all

---

1: **procedure** MSPTYPE(MSP P)
2:    **for all** $(E_x, E_y \in P) \land (x \leq i) \land (y > i) \land (i, y \leq n)$ **do**
3:       **if** $EventType(E_x) == EventType(E_y)$ **then**
4:          **if** $EventType(E_x) == R$ **return R**
5:          **if** $EventType(E_x) == F$ **return F**
6:          **if** $EventType(E_x) == D$ **return D**
7:       **else**
8:          **if** $EventType(E_x) == R$ & $EventType(E_y) == F$ **return RF**
9:          **if** $EventType(E_x) == R$ & $EventType(E_y) == D$ **return RD**
10:          **if** $EventType(E_x) == F$ & $EventType(E_y) == R$ **return FR**
11:          **if** $EventType(E_x) == F$ & $EventType(E_y) == D$ **return FD**
12:          **if** $EventType(E_x) == D$ & $EventType(E_y) == R$ **return DR**
13:          **if** $EventType(E_x) == D$ & $EventType(E_y) == F$ **return DF**
14:       **end if**
15:    **end for**
16: **end procedure**

Fig. 3. Finding MSP Type of a given MSP.

TABLE III
MAXIMAL SEQUENTIAL PATTERN TYPES

| Type | Description |
|------|-------------|
| R | All events must be registry |
| F | All events must be file |
| D | All events must be actions on dll files |
| RF | The MSP has one or more transitions while the first transition is from a registry event to a file event |
| RD | The MSP has one or more transitions while the first transition is from a registry event to a dll event |
| FR | The MSP has one or more transitions while the first transition is from a file event to a registry event |
| FD | The MSP has one or more transitions while the first transition is from a file event to a dll event |
| DR | The MSP has one or more transitions while the first transition is from a dll event to a registry event |
| DF | The MSP has one or more transitions while the first transition is from a dll event to a file event |

---

ransomware sequences ($\gamma$) in a given dataset *D*. For every sequence *S* we can define a Vector of size nine (9) that contains *SR* value of every *MSP* type detected in *MSP Collection MC* within sequence *S* as follow:

$Vector(S)_{MC} = \{(SR_R), (SR_F), (SR_D), (SR_{RF}), (SR_{RD}), (SR_{FR}), (SR_{FD}), (SR_{DR}), (SR_{DF})\}$

*SR* value of every *MSP Type* of a sequence for a given *MC* can be calculated using *CalculateSR* procedure shown in Fig. 4. When vector of all sequences within a sequential dataset *D* using a *MSP Collection MC* is created, we will have a

```
1: procedure CALCULATESR(Sequence S, MSP Collection
    MC, MSPType_Set T)
2:    SR_{P_Total} = 0
3:    for all P ∈ MC do
4:       if P ⊆ S & MSPType(P) == T then
5:          SR_{P_Total} = SR_{P_Total} + (sup_P / γ)
6:       end if
7:    end for
8:    return SR_{P_Total}
9: end procedure
```

Fig. 4. SR calculation algorithm for sequence S.

Vectored Dataset $VD_{D,MC}$.

Moreover, for every sequence *S* we can define a *SuperVector* of *S* as a set of Vectors created using *MSPs* collected from different dataset i.e. $MC_{D_1}, MC_{D_2}, ..., MC_{D_n}$ as follow:
$SuperVector(S) = \{Vector(S)_{MC_1}, Vector(S)_{MC_2}, ..., Vector(S)_{MC_n}\}$ where $MC_1$ to $MC_n$ reflects *MSP Collections* of different datasets. Size of a *SuperVector* with $n$ vectors is calculated by $n \times m$ where $m$ is size of each vector (9 in this research). By calculating *SuperVector* for all sequences within a dataset $D$, we will have Super Vectored Dataset $SVD_D$.

## V. HUNTING FOR EVIL: RANSOMWARE DETECTION

To detect best features (*MSP* types) for classifying ransomware from benign applications we created a dataset $MC_{D_{Total}}$ by combining $D_{Ransomware}$ and $D_{Goodware}$ and then generated a vectored dataset $VD_{D_{Total}, MC_{Ransomware}}$. We then utilized greedy stepwise search method of *CfsSubsetEval* [49] of *Weka3.8.1* with $VD_{D_{Total}, MC_{Ransomware}}$ and found that *MSP Types* of *R* (Registry), *D* (DLLs) and *FD* (File and DLL) may provide best distinction between ransomware and goodware samples (see Fig. 5). As shown in Fig. 5a ransomware applications are tend to conduct a much wider range of Registry activities in compare with gooodware apps. As shown in Fig. 5b, majority of benign applications were conducting similar DLL activities while there were much more variations in ransomware samples DLL events. Ransomware applications are taking a variety of Filesystem to DLL transitions while goodware samples were mainly taking only two specific Filesystem to DLL events transitions (see Fig. 5c).

We have utilized R, D, and FD as features to train four classifiers namely *J48, Random Forest, Bagging*, and *Multi Layer Perception (MLP)* using $VD_{D_{Total}, MC_{Ransomware}}$ and 10-fold cross validation technique for evaluation. As shown in Table IV, all classifiers achieved F-measure of 0.99 with a low false positive rate ($FPR \leq 0.04$). Moreover, similarities between *ROC* curves of different classifiers (see Fig. 6) proves that there is not much difference between performance of different classifiers which is another indication of suitability of our features for classifying ransomware and benign applications. As shown in Fig. 7, *AUC* value for all classifiers is quite high (more than 0.990) while *AUC* value of *Bagging* classifier (0.995) is very close to an optimal prediction. The *MCC* value of all classifiers is more than 0.96 while *Random Forest* and

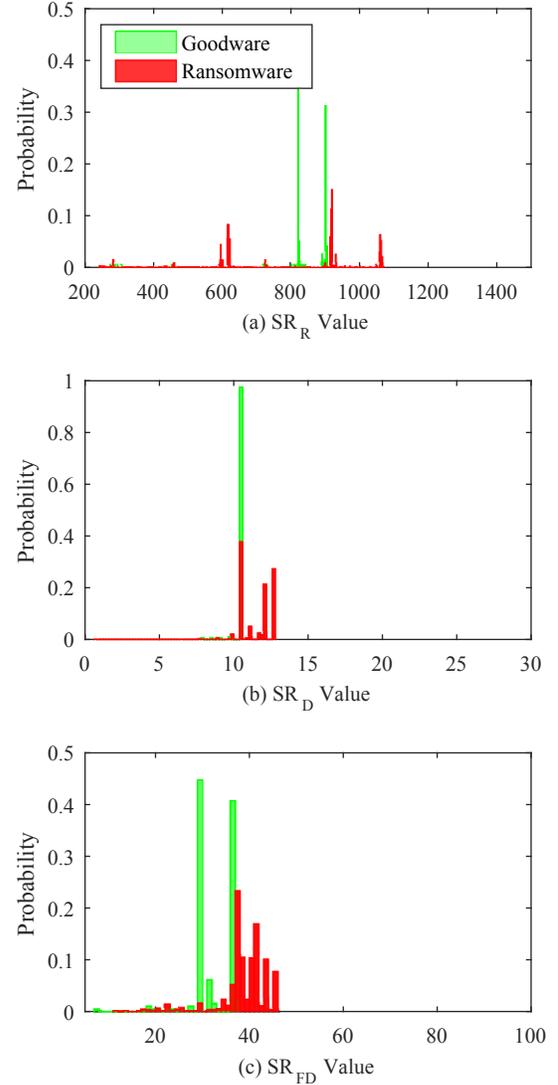

Fig. 5. Histogram of the probability of SR values for ransomware and goodware.

TABLE IV
CLASSIFIERS PERFORMANCE ON $VD_{D_{Total}, MC_{Ransomware}}$

| Classifier | TPR | FPR | F-Measure |
|---|---|---|---|
| J48 | 0.994 | 0.040 | 0.994 |
| Random Forest | 0.993 | 0.040 | 0.993 |
| Bagging | 0.994 | 0.039 | 0.977 |
| MLP | 0.994 | 0.035 | 0.994 |

*Bagging* achieved *MCC* of almost +1 which is very close to a perfect prediction.

To show that we have not over-fitted our classifiers, we tested all classifiers using on $VD_{D_{OF}, MC_{Ransomware}}$. As shown in Table V all classifiers achieved accuracy of 0.994

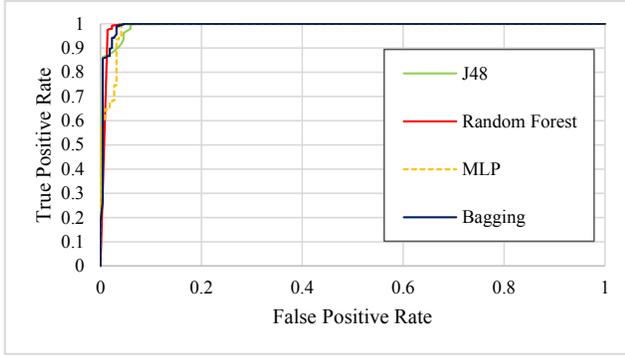

Fig. 6. ROC diagrams for classifiers.

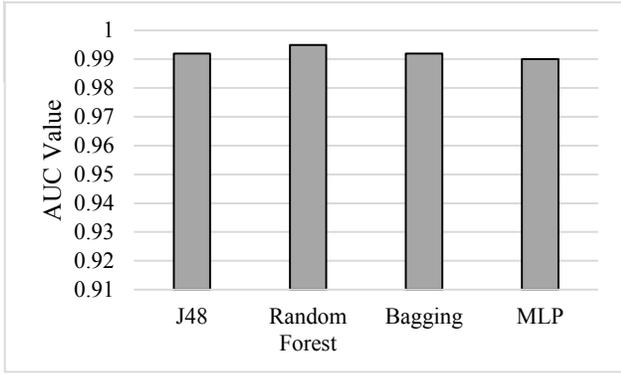

Fig. 7. AUC of classifiers for detecting ransomwares

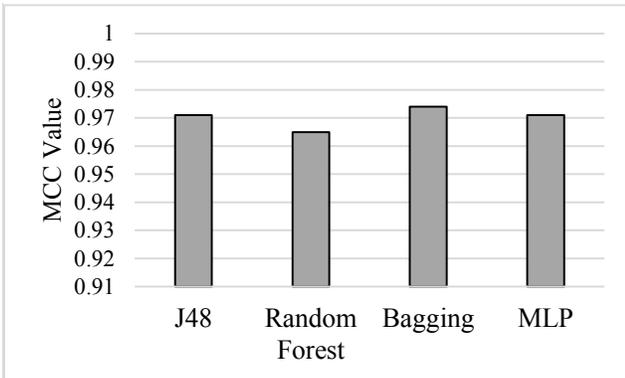

Fig. 8. MCC [24] of classifiers for detecting ransomwares

TABLE V
RESULTS OF CLASSIFIERS ON $VD_{D_{OF}, MC_{Ransomware}}$ FOR DETECTING RANSOMWARE

| Classifier | Accuracy |
|---|---|
| J48 | 0.994 |
| Random Forest | 0.994 |
| Bagging | 0.994 |
| MLP | 0.994 |

in classifying unforeseen ransomware and goodware samples too.

TABLE VI
THE CLASSIFIERS PERFORMANCE ON $SVD_{D_{TotalFamily}}$

| Classifier | TPR | FPR | F-Measure | MCC |
|---|---|---|---|---|
| J48 | 0.981 | 0.006 | 0.981 | 0.974 |
| Random Forest | 0.983 | 0.006 | 0.983 | 0.978 |
| Bagging | 0.980 | 0.007 | 0.980 | 0.974 |
| MLP | 0.980 | 0.007 | 0.980 | 0.973 |

## VI. THREAT INTELLIGENCE: DETECTION OF A RANSOMWARE FAMILY

To investigate performance of classifiers in detection of a ransomware family we have created $D_{TotalFamily}$ dataset which contains all sequences from $D_{Locky}$, $D_{Cerber}$, $D_{TeslaCrypt}$ and $D_{Goodware}$. We then generated $VD_{D_{TotalFamily}, MC_{Locky}}$, $VD_{D_{TotalFamily}, MC_{Cerber}}$ and $VD_{D_{TotalFamily}, MC_{TeslaCrypt}}$ vectored datasets and fed them to *CfsSubsetEval* of *Weka3.8.1*. All in all 13 candidate features were detected for classification of ransomware families as shown in Fig. 9.

Fig. 9 reveals that atomic *Registry MSPs* (activities) are of great importance to differentiate between ransomware families. Scattered values of $SR_R$ for different families in Fig. 9a, 9b and 9c make $SR_R$ as a desirable feature for separation of different families of ransomwares. *Filesystem to DLL* ($SR_{FD}$) is also a useful feature for identifying a ransomware family as shown in Fig. 9a, 9b and 9c. Fig. 9a also shows that *Locky* ransomware samples tend to have more *Registry* activities based on $VD_{D_{TotalFamily}, MC_{Locky}}$ while *Cerber* samples performed more *DLL* activities (see $SR_D$ in Fig. 9b) in compare with other studied families. Moreover, *Locky* samples had the most number of *Registry* to *Filesystem* transitions (based on $VD_{D_{TotalFamily}, MC_{Locky}}$ and based on $VD_{D_{TotalFamily}, MC_{TeslaCrypt}}$ in Fig. 9a and Fig. 9c respectively) and *DLL* to *Filesystem* ($SR_{DF}$) transitions (based on $VD_{D_{TotalFamily}, MC_{Locky}}$ see Fig. 9a and 9b). Finally, *Filesystem* to *Registry* transitions ($SR_{FR}$) are most common within *Cerber* samples (see Fig. 9b).

As detection of ransomware families is a multi-class classification task with four class labels (*Locky, Cerber, TeslaCrypt* and *Goodware*), therefore, we have trained *J48, Random Forest, Bagging* and *MLP* with a multi-class classifier using $SVD_{D_{TotalFamily}}$ dataset with 13 selected features in Fig. 9.

Table VI presents performance of all classifiers obtained from 10-fold cross validation. Obtained minimum weighted average [50] *F-Measure* of 0.983 with $FPR \leq 0.006$ reflects suitability of our features for detecting ransomware samples families. *MCC* values of more than 0.95 for all classifiers also indicate quality of our features in enabling classifiers to provide an almost perfect prediction. Finally, as shown in Table VII our features enabled classifiers to offer an accurate prediction ($\geq 0.965$) even on unforeseen samples ($SVD_{D_{OF}}$).

## VII. CONCLUDING REMARKS

In this paper, by combining sequential pattern mining for feature identification with machine learning classification techniques we could accurately distinguish between ransomware and goodware samples and identify given ransomware families with in first 10 seconds of a ransomware execution. We achieved minimum *F-measure* of 0.994 with minimum *AUC* value of 0.99 in detection of ransomware samples from goodware using Registry (R) events, DLL (D) events and Filesystem to Registry (FD) transitions as features for *J48, Random Forest, Bagging* and *MLP* classifiers. We achieved *F-Measure* of more than 0.98 with *FPR* of less than 0.007 in detection of a given ransomware family using 13 selected features detected in this study. Theoretical implication of this study stems from application of sequence pattern mining to detect frequent features of ransomware applications to build vectored datasets of ransomware logs. Moreover, created dataset of 1624 ransomware samples and 220 benign applications can be used by future researchers to further our understanding of ransomware behavior. Practical implications may include utilization of reported features for differentiating ransomware and benign applications for ransomware threat hunting while features reported for ransomware family classification are great for building intelligence about threat profiles applicable to a given target. In recent times, researchers have proposed the concept of forensic-by-design [51]–[53], and another interesting future research is to extend this forensic-by-design to capturing ransomware detection, mitigation and roll-back. Applying other classification techniques such as fuzzy classification can be considered as a future work of this study. Moreover, utilization of *Stream Data Mining* techniques to reduce ransomware detection time is another interesting extension of this study. Finally, it is interesting to utilize our technique in other emerging domains such mobile malware detection and Internet of Things (IoT) forensics.

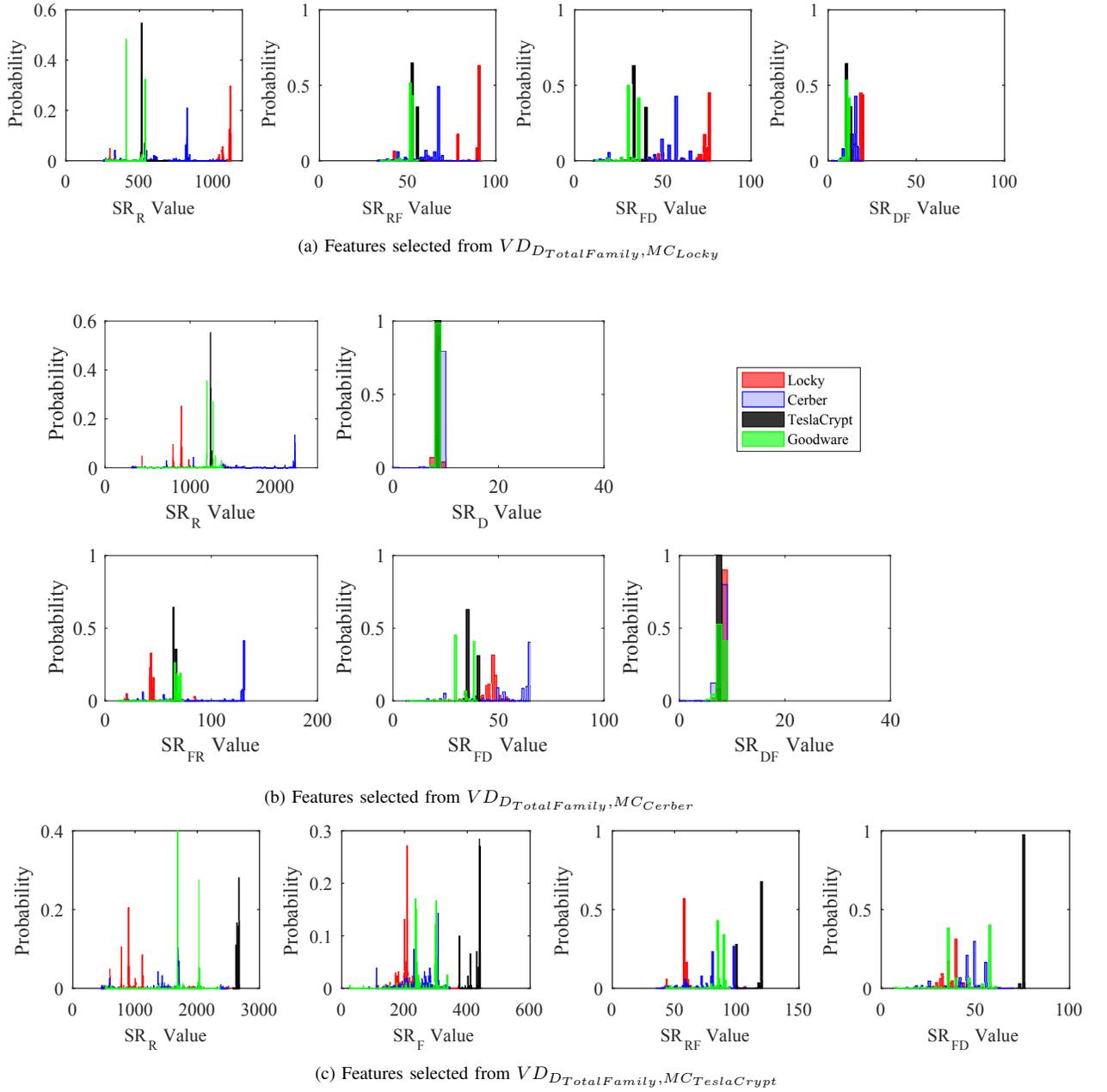

Fig. 9. Histogram of the probability of SR values for ransomware families.

TABLE VII
RESULTS OF CLASSIFIERS ON DATASET $SVD_{D_{OF}}$ FOR DETECTING RANSOMWARE FAMILY

| Classifier | Accuracy |
|---|---|
| J48 | 0.947 |
| Random Forest | 0.965 |
| Bagging | 0.959 |
| MLP | 0.959 |

## Acknowledgment

The authors would like to thank *virustotal.com* for sharing malware samples and giving the capability of scanning files. We also thank *ransomwaretracker.abuse.ch* for their updated and precious information about different families of ransomwares.

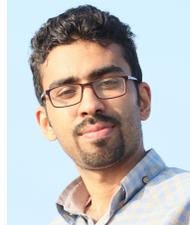

**Sajad Homayoun** is a Ph.D candidate of Computer Networks at Shiraz University of Technology since 2013. He also has a master's degree in Information Technology form K. N. Toosi University of Technology (Khajeh Nasire Toosi) of Tehran and a bachelor's degree in Computer Software. He is currently in charge of security laboratory (SecLab) in Shiraz University of Technology since 2014. His research interests are Cyber Security, Machine Learning Applications in Computer Security and Computer Networks.

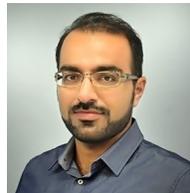

**Ali Dehghantanha** is a Marie-Curie International Incoming Fellow in Cyber Forensics, a fellow of the UK Higher Education Academy (HEA) and an IEEE Sr. member. He has served for many years in a variety of research and industrial positions. Other than Ph.D in Cyber Security he holds several professional certificates such as GXPN, GREM, GCFA, CISM, and CISSP. His main research interests are cyber threat intelligence, threat hunting and digital forensics.

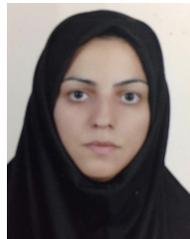

**Marzieh Ahmadzadeh** holds a Ph.D in Computer Science and MSc. In Information Technology, both received from the University of Nottingham, UK, and a first class BSc in Software Engineering received from Isfahan University, Iran. Since September 2006, she has been an assistant professor at the school of computer Engineering and IT, Shiraz University of Technology, where she has supervised more than 20 MSc students whose research area are mostly applied data mining. Being a research assistant at the University of Nottingham and full stack software engineer for 3 years is also part of her work experience. Her research interest includes Data Mining, Data Security, HCI and Computer Science Education.

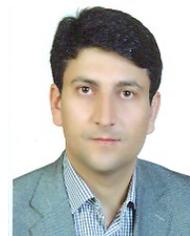

**Sattar Hashemi** received the PhD degree in computer science from Iran University of Science and Technology in conjunction with Monash University, Australia, in 2008. Following academic appointments at Shiraz University, he is currently an associate professor at Electrical and Computer Engineering School, Shiraz University, Shiraz, Iran. His research interests include machine learning, data mining, social networks, data stream mining, game theory, and adversarial learning.

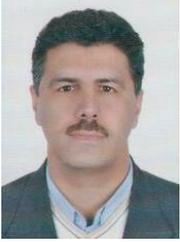

**Raouf Khayami** received his BS degree in computer engineering (hardware systems) from Shiraz University in 1993, the MS degree in artificial intelligence and robotics from the same university in 1996, and the Ph.D degree in software systems from Shiraz University in 2009. He is currently an assistant professor in the Computer Engineering and Information Technology Department, Shiraz University of Technology, Shiraz, Iran, and there, he is the Head of the Department. His research interests include data mining, business intelligence, and enterprise architecture, on which he has published a number of refereed articles, surveys and technical reports in prestigious national and international conferences and journals. He also is active in consulting and industrial projects.